\documentclass[fleqn,usenatbib]{mnras}

\usepackage{graphicx}
\usepackage{amssymb}
\usepackage{amsmath}
\usepackage{enumitem}
\usepackage{natbib}
\usepackage{booktabs}
\usepackage{multirow}

\usepackage{newtxtext,newtxmath}

\usepackage[T1]{fontenc}
\usepackage{ae,aecompl}

\usepackage[usenames,dvipsnames,svgnames,table]{xcolor}

\title[GMC Clumps: Map and Environment]{Northern Galactic Molecular Cloud Clumps in Hi-GAL: Dense Gas Map and Environmental Trends} 

\author[E. Zetterlund et al.]{
Erika Zetterlund,$^{1}$
Jason Glenn,$^{1}$
Erik Rosolowsky$^{2}$
\\
$^{1}$CASA, Department of Astrophysical and Planetary Sciences, University of Colorado 389-UCB, Boulder, CO 80309, USA\\
$^{2}$Deptartment of Physics, University of Alberta, Edmonton, Alberta, Canada\\
}

\date{Accepted XXX. Received YYY; in original form ZZZ}

\pubyear{2018}

\begin{document}
\label{firstpage}
\pagerange{\pageref{firstpage}--\pageref{lastpage}}
\maketitle

\begin{abstract}

In the quest to understand high-mass star formation, it is necessary to understand that from which the high-mass stars will form -- the dense molecular gas clouds and clumps. The \textit{Herschel} infrared GALactic plane survey (Hi-GAL) is a comprehensive survey of thermal dust emission that can be used to characterize the properties and Galactic distribution of molecular gas. We have analyzed the survey maps within the Galactic longitude range $10^\circ < \ell < 56^\circ$ and have transformed these 2D maps into a 3D dense gas map of the Galactic plane using distance probability density functions. This range corresponds to the extent of the Galactic Ring Survey (GRS), which provided the majority of our line-of-sight velocities, and thus kinematic distances. In this section of Hi-GAL, we identify 19,886 clumps, out of which 10,124 have, at minimum, a kinematic distance, and 5,405 have their distances well constrained through Bayesian techniques. Of those with well-constrained distances, clump masses tend to decrease with Galactocentric radius, whereas clump radii are independent of Galactocentric radius.

\end{abstract}


\section{Introduction}

As it is the host of star formation, it is important that we understand the distribution and properties of molecular gas in our Galaxy. While much progress, theoretical and observational, has been made in the study of star formation, understanding the apparent uniformity of the initial mass functions for stars and star clusters \citep[e.g.][]{Kroupa02} requires understanding their precursor material. However, not all molecular gas will form stars.  Rather, it is {\it dense} molecular gas \citep{Gao04} that shows the closest connection to the star formation process. The correlation between the amount of dense molecular gas and the star formation rate appears linear on both global and local scales \citep{Wu07}.

This linear relationship has been the basis for many effective models of star formation, especially in simulations of star formation and galaxy evolution. Current simulations use prescriptive recipes for star formation which are tuned such that the empirical relationship is produced, as opposed to using physically motivated parameter choices. A typical example has star formation turned on at a fixed interstellar medium density threshold (e.g. $10^3$ cm$^{-3}$), at which point $\sim1\%$ of the gas mass is converted into stars \citep[e.g.][]{Guedes11,Kim2014,Dobbs15}. This route is taken because star formation is unresolved at the resolution of entire-galaxy simulations, but nevertheless, it presents a problem when these simulations cannot a priori reproduce the strong winds and star formation inefficiency observed in galaxies \citep{Hopkins14}.

More recently, \citet{Usero15} has suggested that this strictly linear relationship does not completely capture the complexities of star formation efficiency. Instead there is a dependency on galactocentric radius, with efficiency being lower nearer galaxy centres. Properties of the clouds as a whole, such as turbulence, may play a part.

With the presence of dense gas being the dominant factor in star formation efficiency, it is important that we understand how dense gas forms from the diffuse molecular clouds. Dense molecular gas structures can be roughly divided into three nested substructures: clouds, clumps, and cores. These structures increase with density at each level in this hierarchy, with cores being the gravitationally bound precursors to individual stars or simple stellar systems.  Typical radii and densities are seen in Table \ref{tab:ccc}. With \textit{Herschel}, cores are only resolvable within $\sim 1$~kpc of the Sun. Clumps are resolvable within $\sim 7$ kpc. Beyond this, we can only resolve entire clouds \citep{BGPS7}

\begin{table}
\centering
\begin{tabular}{lccc}
\hline
                    Parameter & Cloud    & Clump       & Core        \\ \hline
$R$ {[}pc{]}        & $1-7.5$  & $0.15-1.5$  & $0.015-0.1$ \\
$n$ {[}cm$^{-3}${]} & $50-500$ & $10^3-10^4$ & $10^4-10^5$ \\ \hline
\end{tabular}
\label{tab:ccc}
\caption{Typical radii and densities for molecular clouds, clumps, and cores \citep{Bergin2007}.}
\end{table}

The molecular cloud clump mass function is expected to follow either a power-law or lognormal distribution function, depending on which physical process dominates clump formation. Gravitational collapse of dense structures is expected to produce a power-law distribution \citep[e.g.,][]{Padoan02}, whereas supersonic turbulence is expected to produce a lognormal density distribution \citep[e.g.,][]{Padoan97}. However, supersonic turbulence is not necessary for a lognormal distribution \citep{Tassis10}. As is the case with many such competing theories, it is likely that the answer lies in a combination of the two \citep[e.g.,][]{Hopkins13,Offner14,Basu15}. Determining which of these processes dominates the molecular cloud clump formation is essential for evaluating competing theories of high-mass star formation \citep[e.g.,][ for theoretical and observational descriptions respectively]{Elmegreen85, BGPS13}.

In addition to mass functions, we can also study the physical distribution of the molecular cloud clumps. The radial distribution of GMCs has been investigated in the Milky Way \citep[e.g.,][]{Scoville87,Bronfman00}, and in external galaxies \citep[e.g.,][]{Rosolowsky07,Gratier12,Freeman17}, generally finding a peak at intermediate galactocentric radii. A related question regards whether the GMC mass distribution is dependent on galactocentric radius \citep{Rosolowsky07}, or location in a spiral arm vs. an interarm region \citep{Stark06, Colombo14}. While current simulations are more likely constrain the overall baryonic mass distribution than specifically on the spatial distribution of dense gas, those by \citet{Fujimoto14} found that GMCs have mass distributions whose peaks are unaffected by environment. However, the tails of these distributions show a higher prevalence of larger, more massive clouds nearer to the galactic center.

Historically, the field used CO line surveys to trace the structure of the diffuse molecular gas \citep[e.g.,][]{Scoville75,Solomon87}. However, the interpretation of CO line observations is complicated by a number of issues involving line excitation. In particular, critical densities are temperature dependent, leading to ambiguities in the mass being measured. Additionally, CO does not trace molecular hydrogen at a constant ratio, due to freeze-out and abundance variations. Freeze-out dominates when densities exceed $\sim 3 \times 10^4$ cm$^{-3}$ \citep{Bacmann02}, making it highly relevant to studies of dense molecular cloud clumps. Furthermore, there is an envelope in clouds where H$_2$ self-shields and dust is present, but CO is not \citep{Wolfire10}. Finally, the lowest-J transitions of CO are generally optically thick, complicating mass estimations. Thermal dust continuum surveys provide a complementary view of star forming gas. While such maps do not provide the kinematic information that CO line surveys do, optically thin emission allows us to see through the entire Galactic disk. Furthermore, the dust opacities are known to a factor of two \citep{OH94}, allowing for well-constrained masses. 

In recent years, dust continuum surveys have been performed covering much of the Galactic Plane,  enabling the study of molecular cloud clumps across diverse Galactic environments. These have taken the form of dust continuum surveys of the Galactic plane in the millimeter and submillimeter wavelengths  [BGPS: \citet{Aguirre2011,BGPS9}; ATLASGAL: \citet{Schuller09,Csengeri14,Csengeri16}]. Through these surveys, tens of thousands of molecular cloud clumps and cores have been detected, and their properties can now be analyzed. By extracting these sources and creating a census of molecular cloud structures, theories of star formation and galaxy evolution can be constrained \citep[e.g.,][]{Kennicutt12}. Such studies of physical properties \citep[e.g.,][]{Peretto10,Giannetti13} and mass functions \citep[e.g.,][]{Netterfield09,Gomez14,BGPS12,Wienen15} have already begun, but have not yet produced a consensus answer to questions concerning the evolution of these dense molecular gas structures and their production of a uniform stellar cluster mass function.

However, both BGPS and ATLASGAL suffer from low sensitivity due to being ground-based surveys. A great leap in sensitivity has been made recently with space-based Hi-GAL \citep{Molinari2010,Molinari16b}.  \textit{Herschel Space Observatory's} Hi-GAL survey provides complete coverage of the Galactic Plane in the far-infrared to submillimeter wavebands.
In this paper, we map the portion of Hi-GAL which overlaps with the Galactic Ring Survey (GRS) --- the $^{13}$CO(1-0) survey which provides our line-of-sight velocities --- and investigate the distribution of dense molecular gas across this portion of the Milky Way, which should be representative of the inner Galaxy as a whole.
\citet{Molinari16b} have independently begun cataloging the dense molecular cloud clumps found in the Hi-GAL survey. They have identified and extracted photometry for clumps found in the majority of the inner Galaxy in all five of \textit{Herschel's} photometric bands. This work serves as a complement to the work of \citet{Molinari16b} through the use of a fundamentally different clump identification and distance determination techniques. Hi-GAL has also been used to advance the above studies of physical properties \citep{Elia13} and mass functions \citep{Olmi13,Olmi14}.

Because of the excellent quality of the  {\it Herschel} Hi-GAL survey, we have a new opportunity to explore the trends in molecular gas with respect to galactocentric radius. We build on the previous efforts of \citet{Zetterlund17} which produced a catalog of compact emission structures over the Hi-GAL survey volume. From that work, we apply the distance determination methods used for the BGPS \citep{BGPS8}, yielding a broader survey of molecular cloud clump properties than was previously established.  We then explore the trends in this clump catalog with respect to both heliocentric and galactocentric distances.

\section{Data}

The \textit{Herschel} infrared GALactic plane survey (Hi-GAL) \citep{Molinari2010} observed a $2^\circ$ strip covering the entire $360^\circ$ of longitude of the Galactic plane. Using the SPIRE \citep{Griffin2010} and PACS \citep{Poglitsch2010} instruments aboard the \textit{Herschel Space Observatory} (HSO), this survey observed in wavebands at 70, 160, 250, 350, and 500 $\mu$m. SPIRE (250, 350, and 500 $\mu$m) observed thermal dust emission which is concentrated in dense molecular gas clumps. Like SPIRE, PACS (70 and 160 $\mu$m) observed thermal dust emission from dense gas. However, PACS observed extended warm structures to a greater extent than SPIRE. Owing to the low optical depth of thermal dust in the submillimeter continuum, Hi-GAL allows us to view dense molecular gas throughout the entire Galactic disk.

In this paper, we analyzed Hi-GAL regions with Galactic longitudes $10^\circ < \ell < 56^\circ$. This range was chosen as corresponding to the limits of the $^{13}\text{CO}$ Galactic Ring Survey \citep[GRS;][]{Jackson06} ($18^\circ < \ell < 56^\circ$), plus a few extra degrees near the Galactic center. The expanded range allows us to match the coverage of the CO High-Resolution Survey (COHRS), which we will integrate into our analysis in a future paper. GRS probes dense gas in the Galactic Plane using the $J=1-0$ transition in $^{13}$CO for the velocity range --5 to 85 km s$^{-1}$. We have pointed observations of the $J=3-2$ rotational transitions of HCO$^+$ and N$_2$H$^+$ made for BGPS clumps from \citet{Shirley13} to provide velocities for this additional longitude not covered by GRS.

\section{Method}

Following \citet{Zetterlund17}, we apply an angular filter to the 500-$\mu$m Hi-GAL maps in $2^\circ \times 2^\circ$ tiles in order to remove large-scale flux density originating from cirrus clouds. We produce the maps using the pipeline {\sc Unimap} \citep{Piazzo2012,Piazzo2015a,Piazzo2015b}.  We then apply a high-pass Gaussian filter with $\sigma = 3\arcmin$ to the maps. The orthogonal scanning used to produce Hi-GAL observations leaves each map with ragged edges not covered by both scan directions. Post-filtering, we mask off these edges, as well as a small number of pixels at high absolute Galactic latitudes. At these pixels, the large-scale flux density is negative, leading to the filter adding flux density as opposed to removing it.

We use {\sc Bolocat}, a seeded watershed algorithm tool developed for BGPS \citep{BGPS2}, to identify sources in the processed maps. {\sc Bolocat} identifies areas of significant emission and divides those areas into separate objects based on the difference in emission levels between local maxima and the saddle points between them. The thresholds to be considered significant emission and for dividing objects are based on the noise level of the map. Because Hi-GAL is confusion limited, the noise level of each map cannot usefully be defined by pixel statistics in dark areas of the map. Rather, we define the noise level by the distribution of flux density in the map as a whole. We take the flux density of all the positive flux density pixels in each map and fit their distribution to an exponential function. The various thresholds are set to values proportional to the scale factor of this fit. This method was chosen as it identifies objects consistent with what we would visually expect, over the large range of flux density levels found in maps at various longitudes \citep[see][]{Zetterlund17}.

With this clump catalog, we then determine clump distances using the method developed for BGPS by \citet{BGPS8} and \citet{BGPS12}. Most of the clump distances are based on their measured radial velocities; the distance ambiguity is resolved probabilistically. We determine the line-of-sight velocity with respect to the LSR using two methods. Many individual bright features have their velocities already determined through a suite of dense gas spectroscopy \citep[e.g.,][]{Shirley13}.  Targets without a specific dense gas measurement have their velocities established through morphologically matching the emission in the dust continuum to the emission from $^{13}$CO(1-0) data in the BU-Galactic Ring Survey \citep{Jackson06}. This is done by subtracting an off-source spectrum from an on-source spectrum for each clump and identifying a single peak in the resulting spectrum. The off-source spectrum is extracted by creating a rind around the source, excluding any pixels which belong to another clump. This method works remarkably well in both crowded and sparse environments, identifying a single velocity in approximately 50\% of cases.

To determine whether each clump is located at the kinematic distance on the near or the far side of the Galactic center, we employ probabilistic methods in a Bayesian framework. A distance probability density function (DPDF) is calculated for each clump. The posterior DPDF is found by,
\begin{equation}
	\text{DPDF} = \mathcal{L}(d_\odot | \ell,b,v_\text{LSR}) \prod_i P_i (d_\odot | \ell,b),
\label{eq:dpdf}
\end{equation}
where $\mathcal{L}(d_\odot | \ell,b,v_\text{LSR})$ is the kinematic distance likelihood and the $P_i (d_\odot | \ell,b)$ are various Bayesian priors, which combine multiplicatively to create the posterior probability distribution for distance along the line of sight, often strongly favoring one $v_\text{LSR}$ peak. The likelihood function is calculated by matching the line-of-sight velocities to the Galactic rotation curve from the BeSSeL survey \citep{Reid14} and equally weighting the near and far distances. For choosing between kinematic distances, the most powerful prior uses 8 $\mu$m absorption to place clumps with strong absorption features at the near kinematic distance. This method produces a continuous probability as a function of distance, not simply a discrimination between near and far locations. The methods of \citet{BGPS8, BGPS12} use several other associations to constrain the DPDFs. These additional priors include using the small scale height of molecular hydrogen to constrain clumps found at high Galactic latitudes to be at their near kinematic distances. For a small number of objects, we have robust trigonometric maser parallaxes from the BeSSeL survey. These clumps' $(\ell,b,v_\mathrm{LSR})$ coordinates are matched to 6-dimensional (position, velocity) maser association volumes. Because these geometric distances are independent of the Galactic rotation curve and assumptions about the structure of the Galactic Plane, they offer the most robust of the distance measurements. Once DPDFs are calculated, we establish clump properties from Monte Carlo simulations over the distance distribution, which provides robust uncertainties for cloud clump properties.

Distances establish clump masses and radii, which are central to this study. Physical radii are given as $R = \theta_R d_\odot$, where $\theta_R$ is angular radius and $d_\odot$ is heliocentric distance. Our mass estimate uses the high-pass filtered flux densities at 500 $\mu$m. The conversion from flux density to mass is 
\begin{equation}
	M = \frac{\Gamma d_\odot^2}{\kappa_{500} B_{500}(T)} S_{500},
\label{eq:mass}
\end{equation}
where $\Gamma=100$ is the gas-to-dust mass ratio, $\kappa=5.04$ cm$^2$g$^{-1}$ is the opacity at 500 $\mu$m \citep{OH94}, $B_{500}(T)$ is the Planck function at 500 $\mu$m, $d_\odot$ is the heliocentric distance, and $S_{500}$ is the clump-integrated flux density in the SPIRE PLW (500 $\mu$m) band. \citet{Battersby11}  used pixel-by-pixel modified blackbody fits of Hi-GAL data to determine that mid-infrared-dark molecular cloud clumps generally span the temperature range 15 K $\lesssim T \lesssim$ 25 K. Furthermore, \citet{BGPS7} found an NH$_3$ gas kinetic temperature of $\langle T_K \rangle = 17.4\pm5.5$ K for a sample of 199 BGPS sources. Following these findings, we adopt a universal clump temperature of 20 K for this study.

While using a single {\it Herschel} band and a single representative temperature is not ideal, cirrus emission in the shorter wavelength bands make it impractical to use spectral energy distributions to extract individual clump temperatures. We therefore choose the 500 $\mu$m band, the longest of the {\it Herschel} bands, to calculate masses. This band is as far as we can reach into the Rayleigh-Jeans limit, and thus as close to a $M \propto T^{-1}$ as is available. In this waveband, temperature difference of 5 K at a temperature of 20 K corresponds with a roughly 50\% difference in mass.

\section{Results}

\begin{figure}
\begin{center}
	\includegraphics[width=\columnwidth]{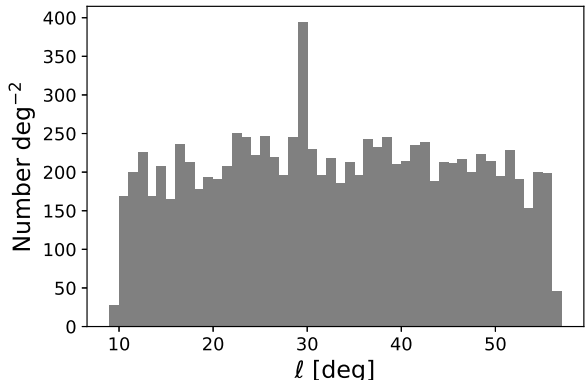}
\caption{The distribution of sources with respect to Galactic longitude. The significant increase at $\ell \sim 30^{\circ}$ shows the enhancement of source number and emission looking through the molecular ring toward the end of the Galactic bar.}
\label{fig:l_dist}
\end{center}
\end{figure}

We identified 19\,886 sources in total, 10\,124 of which have some distance information, and 5\,405 of which have well-constrained distances. The sources are spread evenly across the included Galactic longitudes, with a tendency towards lower flux densities at greater Galactic longitudes. Figure \ref{fig:l_dist} shows the relatively constant number of clumps at each Galactic longitude, with a noticeable spike at $\ell=30^\circ$, due to the end of the Galactic bar. This constancy is caused by Hi-GAL being confusion-limited, as noted in Section 3.

Clumps were considered to have ``well-constrained" distances if their DPDFs had a $\text{FW}_{68} \le 2.3$ kpc. That is, the isoprobability confidence region surrounding the distance of maximum likelihood which contains 68.3\% of the total integrated probability has a full width less than 2.3 kpc. This was adopted from \citet{BGPS8}, which found an empirical cut-off at this point such that clumps with $\text{FW}_{68} \le 2.3$ kpc either had 78\% of the total integrated probability within the most likely kinematic peak, or were located at the tangent distance. Clumps with kinematic distances, regardless of whether the posterior DPDF met the well-constrained criterion, are considered to have some distance information.

\subsection{Malmquist Bias}

\begin{figure}
\begin{center}
	\includegraphics[width=\columnwidth]{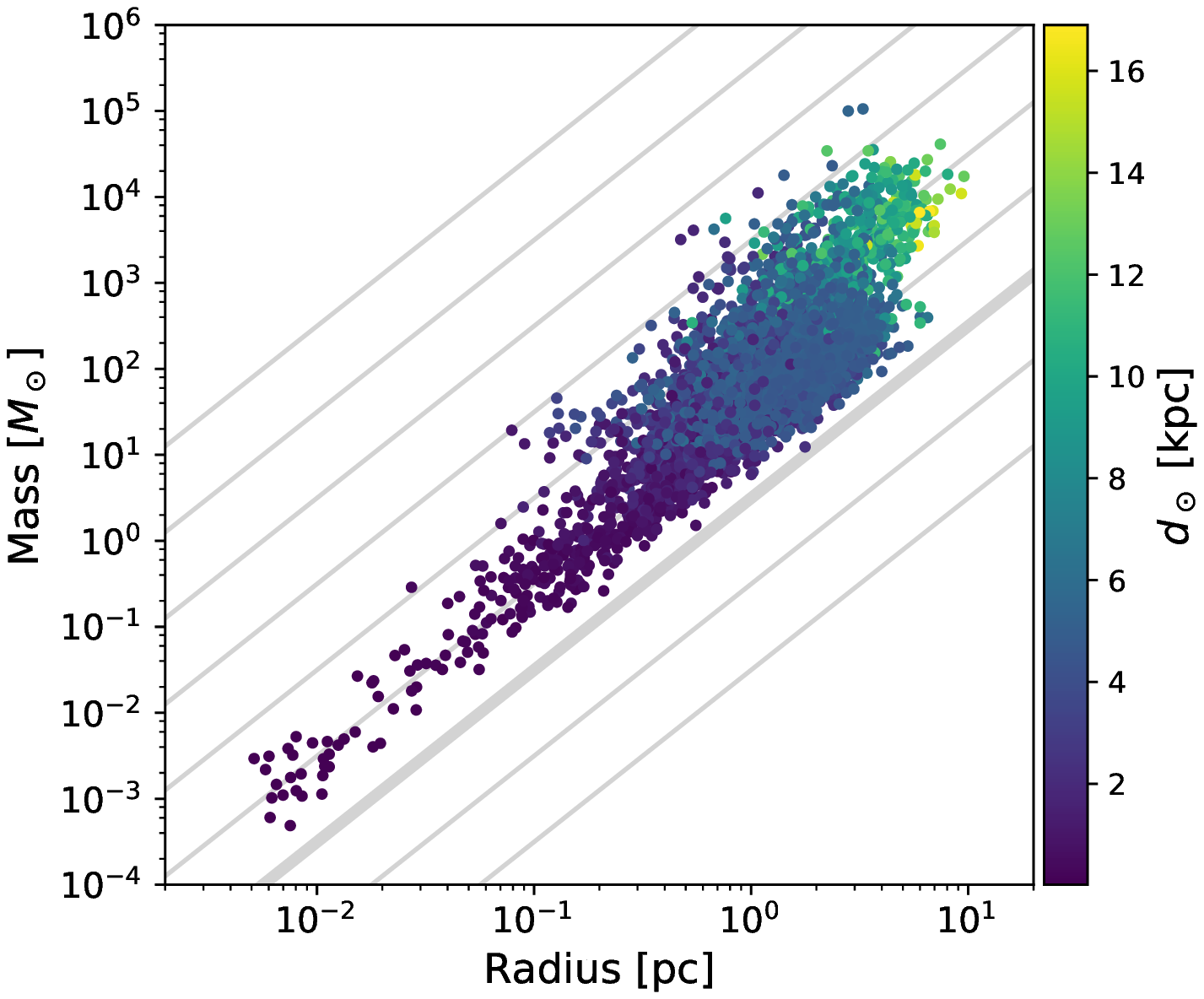}
	\includegraphics[width=\columnwidth]{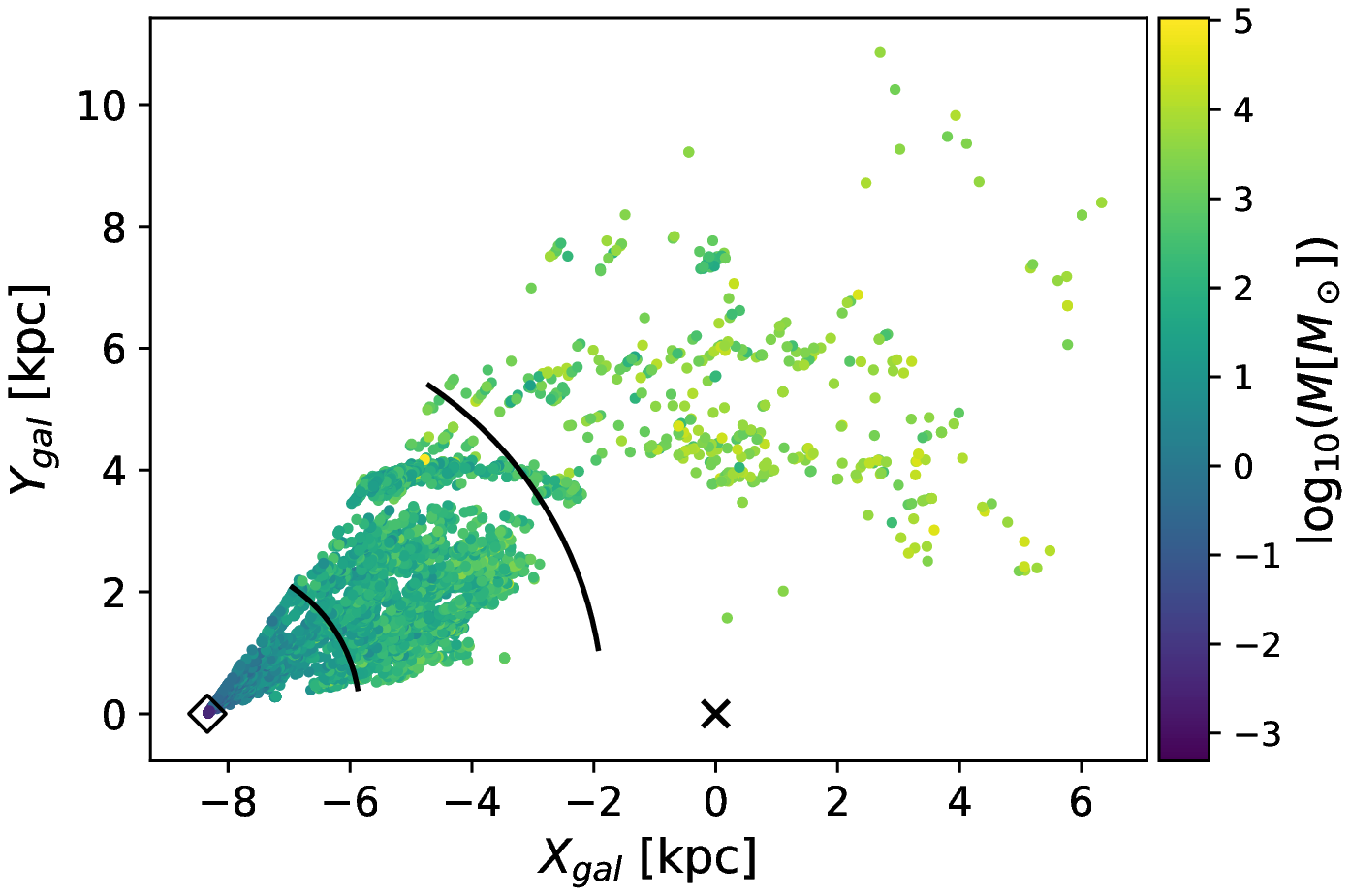}
\caption{\textit{Top}: Mass-Radius diagram for all clumps with well-constrained distances, with colours corresponding to heliocentric distances. Lines of constant surface density are shown in grey. The bold line is $\Sigma=1$ M$_\odot$~pc$^{-2}$, and lines are separated by a factor of ten. \textit{Bottom}: Map of the same clumps, with colours corresponding to clump masses. The solar position and Galactic center are shown as a diamond and an $\times$, respectively. The effects of the Malmquist bias are apparent in both panels in the lack of sources with low mass at large distances. The black arcs indicate the bounds of our minumum-bias sample region, as determined from Figure \ref{fig:lognorms}.}
\label{fig:malmquist}
\end{center}
\end{figure}

Due to the nature of the Hi-GAL survey and our subsequent angular filtering, we are subject to a form of Malmquist bias, identifying only larger, more massive objects at greater distances and smaller objects at nearer distances. This is demonstrated in Figure \ref{fig:malmquist}. The top panel shows a Mass-Radius diagram, coloured by heliocentric distance. There is a clear gradient with distance, showing nearby clumps having low masses and small radii, and distant clumps having high masses and larger radii.  We attribute this relationship to a selection effect which is a product of our angular filtering process and the brightness levels that are required to be recognized as individual objects in our filtered maps. However, for larger clumps ($R>0.3$~pc) there is a significant spread in the distances and surface densities probed.  Over a four order-of-magnitude range in mass, there is a factor of $\sim$10 spread in surface density. We measure a slope of $2.08\pm0.02$ (where the uncertainty given is the standard deviation on the mean) in the mass-radius relationship, which is likely simply an effect of our surface brightness limit.

The lower panel is a face-on map of the Galactic disk showing the positions of the clumps having well-constrained distances, coloured by clump mass. Again there is a clear gradient with less massive clumps near to our position, and more massive clumps farther away.

\begin{figure}
\begin{center}
	\includegraphics[width=\columnwidth]{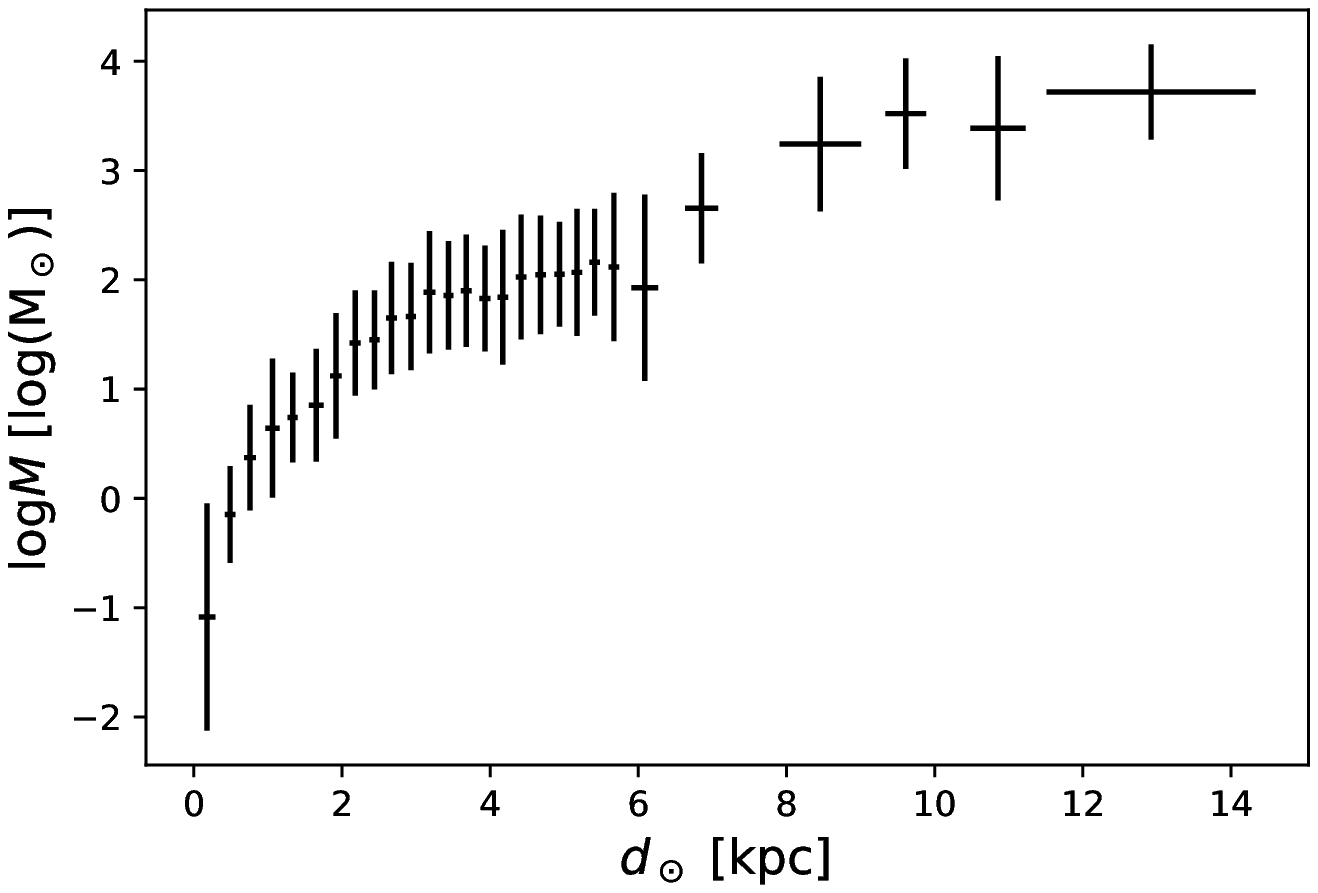}
	\includegraphics[width=\columnwidth]{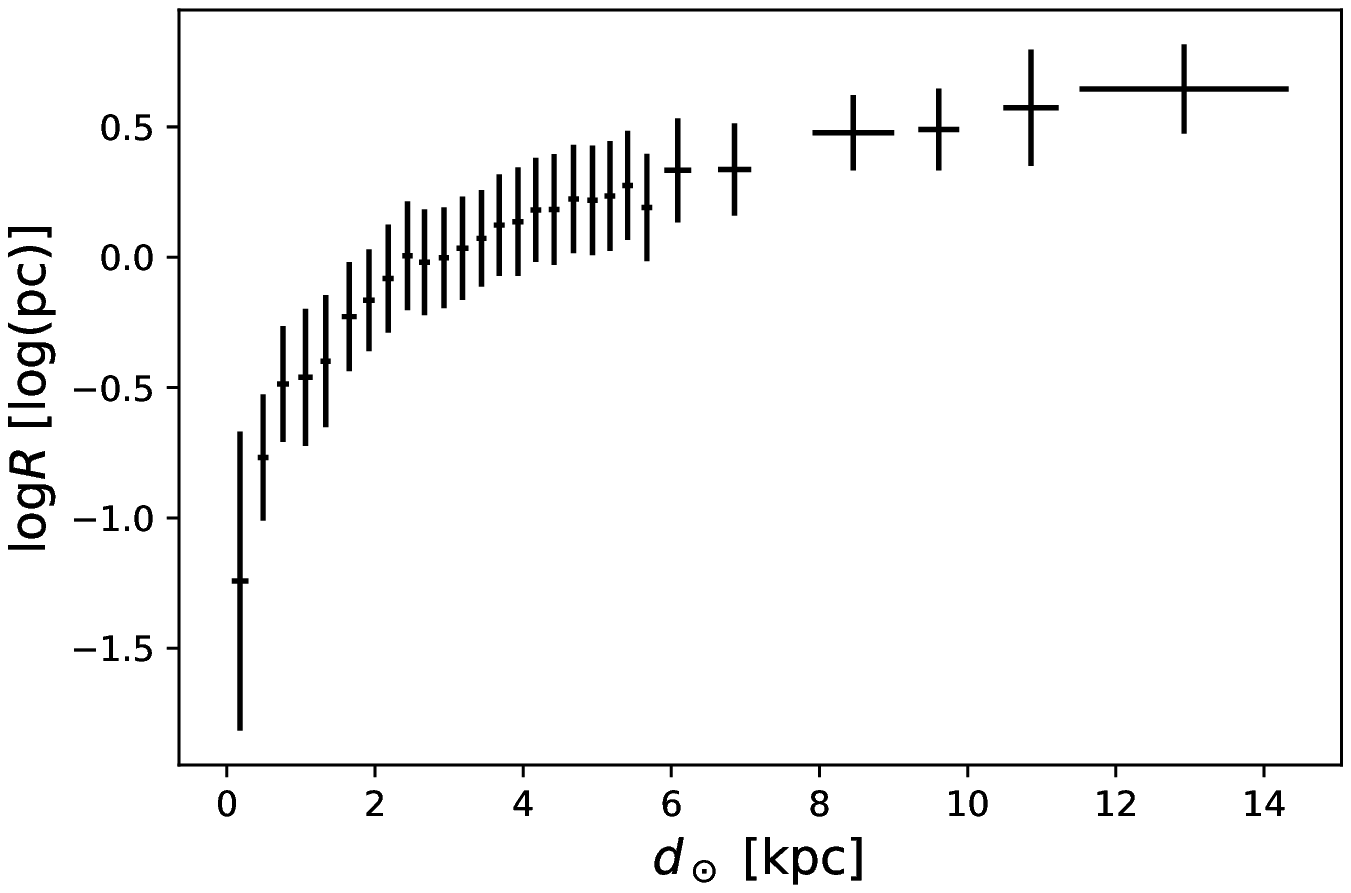}
\caption{Mass and radius trends with heliocentric distance, more clearly shown. \textit{Top}: Clumps separated by heliocentric distance into bins containing a minimum of 100 points and with a minimum width of 0.25 kpc. The mass distribution in each bin is fit to a log-normal, and the mode is plotted. Error bars represent the standard deviation on each axis. \textit{Bottom}: As per the top panel, except for the radius distributions.}
\label{fig:lognorms}
\end{center}
\end{figure}

This bias can be seen more quantitatively in Figure \ref{fig:lognorms}. For each plot, clumps were separated by heliocentric distance into bins containing a minimum of 100 points and with a minimum width of 0.25 kpc. The mass and radius distributions in each bin were fit to lognormal functions. The modes of the distributions are plotted, along with standard deviations in both mass/radius and distance. Over the range of 13 kpc, clump radii increase by more than an order of magnitude, and masses increase by roughly 4 orders of magnitude. However, most of the survey sources are found to lie in the range from $2.5 < d_{\odot}/\mathrm{kpc} < 6.5$. In this region, the distributions of mass as a function of distance do not show significant change and the effects of the Malmquist bias are minimized.  For this part of the survey, we can probe a range of structures at different {\it galactocentric} radii, thereby assessing the evolution of the clump population with environment in the Galaxy.

\subsection{Trends with Galactocentric Radius}

Here, we select objects from the catalog with a heliocentric distance of 2.5 kpc -- 6.5 kpc, where the distributions of clump mass and radius are minimally varying, as per Figure \ref{fig:lognorms} (top). In this range, the effects of the Malmquist bias are significantly reduced, and it becomes possible to probe trends with galactocentric radius.


\begin{figure}
\begin{center}
	\includegraphics[width=\columnwidth]{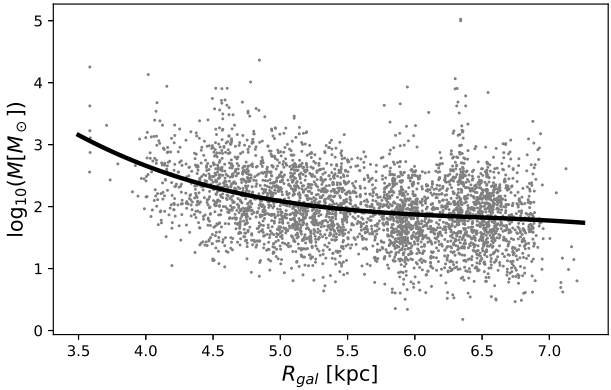}
\caption{Clumps masses plotted against Galactocentric radius for those clumps in the heliocentric distance slice from 2.5 kpc -- 6.5 kpc, where the Malmquist bias is minimized. The line shows a third degree polynomial fit, which represents the typical trend seen in the data and is meant merely as a guide for the eye, not a physical model.  On average, our methods find higher mass clumps in the inner Galaxy compared to the outer Galaxy.}
\label{fig:MRgal}
\end{center}
\end{figure}

\subsubsection{Mass Distribution with Radius}

Figure \ref{fig:MRgal} shows the trend of clump mass with Galactocentric radius within this slice of heliocentric distance.  In this region, the clump mass appears to decrease moving outward from the molecular ring.  The scatter in the cloud masses is large, but the median cloud mass decreases by nearly an order of magnitude over this range of radius.  While source crowding may prevent the small clumps from being seen toward the molecular ring ($M<10^{2}~M_{\odot}$), the upper envelope of the trend indicates that higher mass clumps appear to be found in the molecule-rich region near $R_{\mathrm{gal}}\sim 4.5$~kpc. 

\begin{figure}
\begin{center}
	\includegraphics[width=\columnwidth]{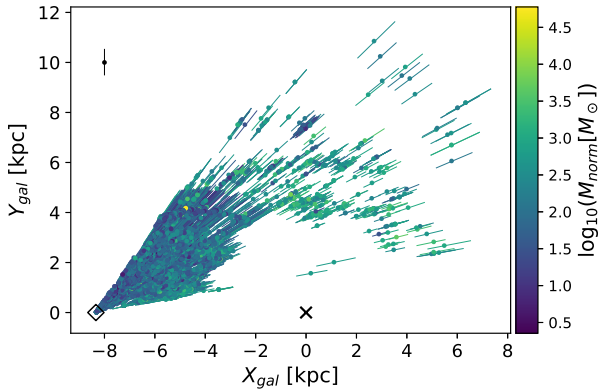}
\caption{Map of clumps with well-constrained distances, as in Figure \ref{fig:malmquist}. colours correspond to clump masses which are normalized to the mass they would have if placed at the mean heliocentric distance of the clumps in our sample. Error bars show the uncertainty in heliocentric distance, and the black point shows representative error bars for clumps with $d_\odot<5$kpc. The solar position and Galactic center are shown as a diamond and an $\times$ respectively.}
\label{fig:map_Mn}
\end{center}
\end{figure}

\begin{figure}
\begin{center}
    \includegraphics[width=\columnwidth]{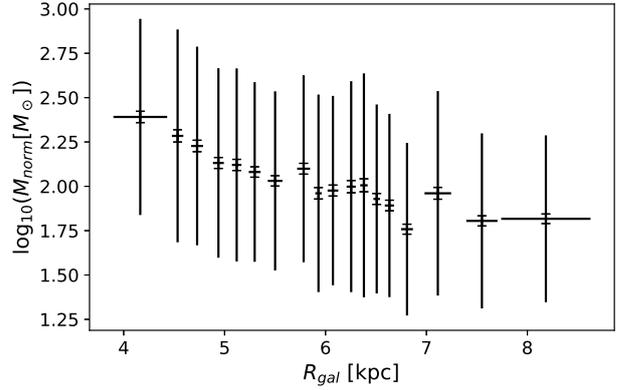}
	\includegraphics[width=\columnwidth]{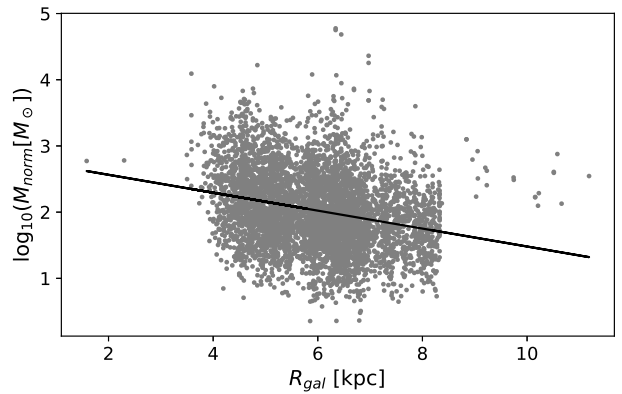}
\caption{Normalized mass is negatively correlated with Galactocentric radius. \textit{Top}: Clumps separated by Galactocentric radius into bins of N=300. Full error bars show standard deviations on both axes. Error bar caps show standard deviation on the mean for mass. \textit{Bottom}: The same data un-binned and fit to a linear function. While there is a large spread in masses, there is also a clear negative trend. Note that the axis ranges in the two panels are different.}
\label{fig:MnRgal}
\end{center}
\end{figure}

\subsubsection{Normalized Masses}
As an alternative strategy to mitigate the Malmquist bias, we can examine the clump population with distance-independent methods. Here, we consider the ``normalized'' clump mass, created by generating clump mass as if all the sources were located at the mean distance of the survey: $d_\odot = 4.12$ kpc. That is, Equation \ref{eq:mass} is used as for true masses, but with $d_\odot = 4.12$ kpc for all clumps. This is tantamount to examining the distribution of clump-integrated flux densities, except we place those quantities on a common physical scale to guide our interpretation of the results. Figure  \ref{fig:map_Mn} presents a normalized version of Figure \ref{fig:malmquist} (bottom), now coloured by normalized mass, and with the addition of heliocentric distance error bars. The bias seen in Figure \ref{fig:malmquist} is no longer present.

Using normalized masses allows us see if the negative trend in mass with Galactocentric radius from Figure \ref{fig:MRgal} remains when the Malmquist bias is minimized using a second independent strategy. We show this trend of normalized mass ($\propto$ clump-integrated flux density) in Figure \ref{fig:MnRgal}. Here we show that, on average, sources in the inner Galaxy have higher characteristic flux densities than those at large galactocentric radius. On the top, clumps were separated into bins of N=300 by $R_\text{gal}$. On the bottom, the unbinned data were fit to a linear function. In both, a clear negative trend is seen, despite a wide spread in individual data points. Overall the flux densities of clumps are decreasing with galactocentric radius, consistent with the trend in the mass determinations seen in Figure \ref{fig:MRgal}.

\subsection{Surface Densities of Clumps}
\begin{figure}
\begin{center}
	\includegraphics[width=\columnwidth]{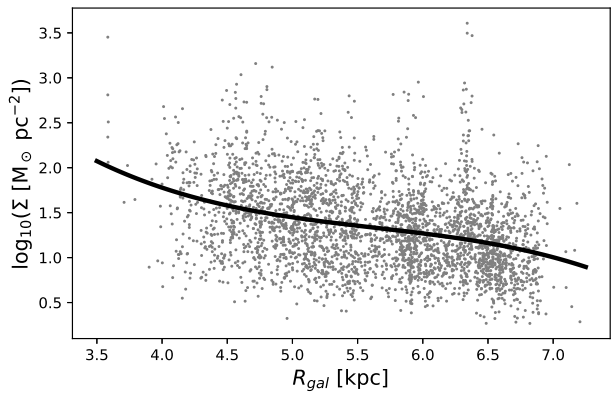}
\caption{Clump surface density plotted against Galactocentric radius for those clumps in the heliocentric distance slice from 2.5 kpc -- 6.5 kpc, where the Malmquist bias is minimized. The line represents a third degree polynomial fit, which represents the typical trend seen in the data and is meant merely as a guide for the eye, not a physical model.}
\label{fig:SigmaRgal}
\end{center}
\end{figure}

Figure \ref{fig:SigmaRgal} shows the mean surface densities of clumps in the analysis. This property is particularly useful to study since it is proportional to the surface brightness of the emission and therefore independent of distance to the source. The typical (median) surface density of one of the clumps in this study is 20$~M_{\odot}~\mathrm{pc}^{-2}$, similar to that seen other wide-area continuum studies \citep[e.g.][]{Kauffmann10}. This average surface density is lower than the empirical minimum surface density for hosting massive star formation \citep[typically about $250\mathrm{~M_{\odot}~pc^{-2}}$,][]{Kauffmann10}, but there are ample numbers of high surface density clumps in this sample volume. Furthermore, those clumps that don't appear star-forming at Hi-GAL's resolution may meet the surface density condition at higher resolutions.

The typical surface density of the clump appears to show an increase in the inner part of the Galaxy, coincident with the molecular ring. At least part of this effect arises from the difficulty of identifying low surface brightness sources against the bright background.  However, the trend is also visible for $R_\mathrm{gal}$ in the vicinity of 4.5 kpc, where crowding is less significant. The increase in typical surface densities likely reflects the increasing pressure in the molecular medium changing its underlying structure \citep[e.g.,][]{Ostriker10}.

\subsection{Mass Distributions}

We explore the mass distribution of the clumps identified in the survey following the approach of \citet[][hereafter F17]{Freeman17}. We restrict the  cloud sample to heliocentric distances in the survey for which the Malmquist bias is minimized (2.5 to 6.5 kpc).  For this sample, we complete a fit to the full population and then divide the sample into 6 bins of galactocentric radius with an equal number of clouds in each bin (126). For these binned mass distributions, we then fit the complementary cumulative distribution functions of the mass spectrum:
\begin{equation} 
\mathrm{CCDF} = 1-\frac{N(>M)}{N_{\mathrm{tot}}}=1-\frac{1}{N_{\mathrm{tot}}}\int^\infty_M \frac{dN}{dM'}dM'
\end{equation}
with a functional form
\begin{equation} 
\label{eq:powerlaw}
\frac{dN}{dM} = M^{\beta} \exp\left(-\frac{M}{M_c}\right).
\end{equation}
We consider two main models for this form of the mass distribution: a pure power-law (PL) where $M_c\to \infty$ and a truncated power-law (TPL) where the truncation mass ($M_c$) is determined through fit to the data. To analyze the mass distributions, we fit the CCDF using an Anderson-Darling goodness-of-fit metric as described in F17. To determine parameter uncertainties, we use the {\sc emcee} \citep{emcee} package to sample the posterior probability density function of the fit parameters and report the errors as the difference between the median values and the 16th and 84th percentile of the value, which would correspond to $-1\sigma$ and $+1\sigma$ respectively if the parameter were normally distributed.  For each mass distribution, we restrict the fit to $M>400~M_{\odot}$ to study the population of massive clumps.  Below this threshold, the distributions show some evidence for deviation from this model, but it is unclear as to whether this represents the effects of blending in the sample at far distances or a change in the functional form of the distribution (i.e., to the log-normal part of a Pareto-log-normal distribution \citealt{Reed04} such as that suggested by \citealt{Brunt15, Basu15}).  The large positive uncertainties in the truncation masses reflect a covariance in the fit parameters: high truncation masses can be compensated for in fitting by a steeper (more negative) power-law index, which is seen in the negative tail of the index uncertainties.

\begin{table*}
\begin{tabular}{r r r r r r r r}
\hline
 & \multicolumn{6}{c}{Radial Bin (kpc)} \\
 Property & 3.6$-$7.0 & 3.6$-$4.5 & 4.5$-$4.8 & 4.8$-$5.1 & 5.1$-$5.5 &  5.5$-$6.3 & 6.3$-$7.0 \\
\hline
Number of Clouds &  756 &  126 &  126 &  126 &  126 &  126 &  126  \\
$M_{\mathrm{max}}$ ($10^{3}~M_{\odot}$) &  97.4 & 17.3 & 10.8 & 23.7 &  2.7 &  4.6 & 97.4 \\
$\langle M \rangle_5$ ($10^3 M_{\odot}$)  & 28.0 & 7.7 &  7.6 &  4.6 &  2.3 &  2.8 & 16.0 \\
$\beta_{\mathrm{PL}}$ &  $-2.27^{+0.05}_{-0.05}$ & $-1.96^{+0.09}_{-0.12}$ & $-1.94^{+0.09}_{-0.11}$ & $-2.29^{+0.12}_{-0.15}$ & $-2.50^{+0.14}_{-0.18}$ & $-3.09^{+0.19}_{-0.24}$ & $-2.27^{+0.11}_{-0.12}$ \\
$\beta_{\mathrm{TPL}}$ & $-2.23^{+0.05}_{-0.07}$ & $-1.59^{+0.05}_{-0.38}$ & $-1.55^{+0.10}_{-0.39}$ & $-2.19^{+0.13}_{-0.18}$ & $-0.57^{+0.57}_{-1.82}$ & $\cdots$ &  $\cdots$  \\
$M_{c,\mathrm{TPL}}$ ($10^3 M_{\odot}$) & $54.8^{+187.2}_{- 0.1}$ & $ 6.1^{+121.9}_{- 0.1}$ & $ 5.9^{+116.5}_{- 0.1}$ & $19.8^{+177.2}_{- 0.1}$ & $ 0.5^{+19.1}_{- 0.1}$ & $\cdots$ & $\cdots$  \\
$\log_{10} p_{\mathrm{PL}}$ &  $-0.51$ & $-0.97$ & $-1.04$ & $-0.46$ & $-1.69$ & $-0.24$ & $-1.48$  \\
$\log_{10} p_{\mathrm{TPL}}$  &   $-0.40$ &  $-0.32$ & $-0.41$ & $-0.42$ & $-0.23$ & $-\infty$ & $-\infty$  \\
\hline
\end{tabular}
   \caption{\label{table:properties} The properties of clump mass distributions for the whole sample and in bins of galactocentric radius with equal numbers of clumps.  The table reports the maximum mass in each bin ($M_{\mathrm{max}}$), the geometric mean of the five most massive clouds $\langle M \rangle_5$, the indices of the power-law (PL) and truncated power-law fits ($\beta_{\mathrm{PL}}$  and $\beta_{\mathrm{TPL}}$ ), the truncation mass of the TPL fit ($M_{c,\mathrm{TPL}}$) and the log-probabilities of the most credible models for each type $(\log_{10} p)$.}
\end{table*}

\begin{figure}
    \centering
    \includegraphics[width=\columnwidth]{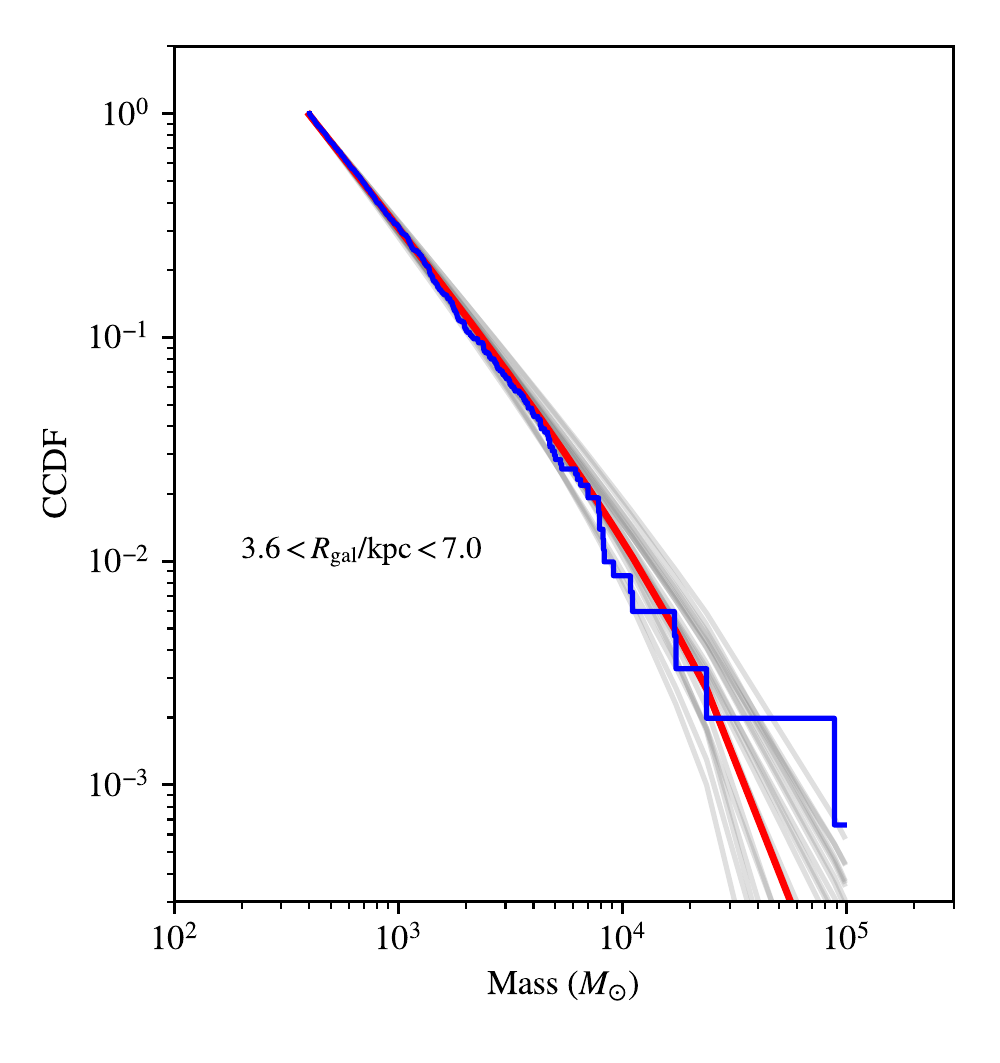}
    \caption{Mass distribution for clumps in our minimal-bias sample.  The blue data show the complementary cumulative distribution function (CCDF) for clump mass and the red curve shows the truncated power-law fit to the CCDF. Grey curves show comparable models drawn from the posterior distributions of the fit parameters.  The overall clump mass distribution function shows good agreement with a power-law model with index of $\beta=-2.27\pm0.05$ over the mass range with only weak evidence for a truncation at the high-mass end.}
    \label{fig:mspecall}
\end{figure}

\begin{figure*}
    \centering
    \includegraphics[width=0.9\textwidth]{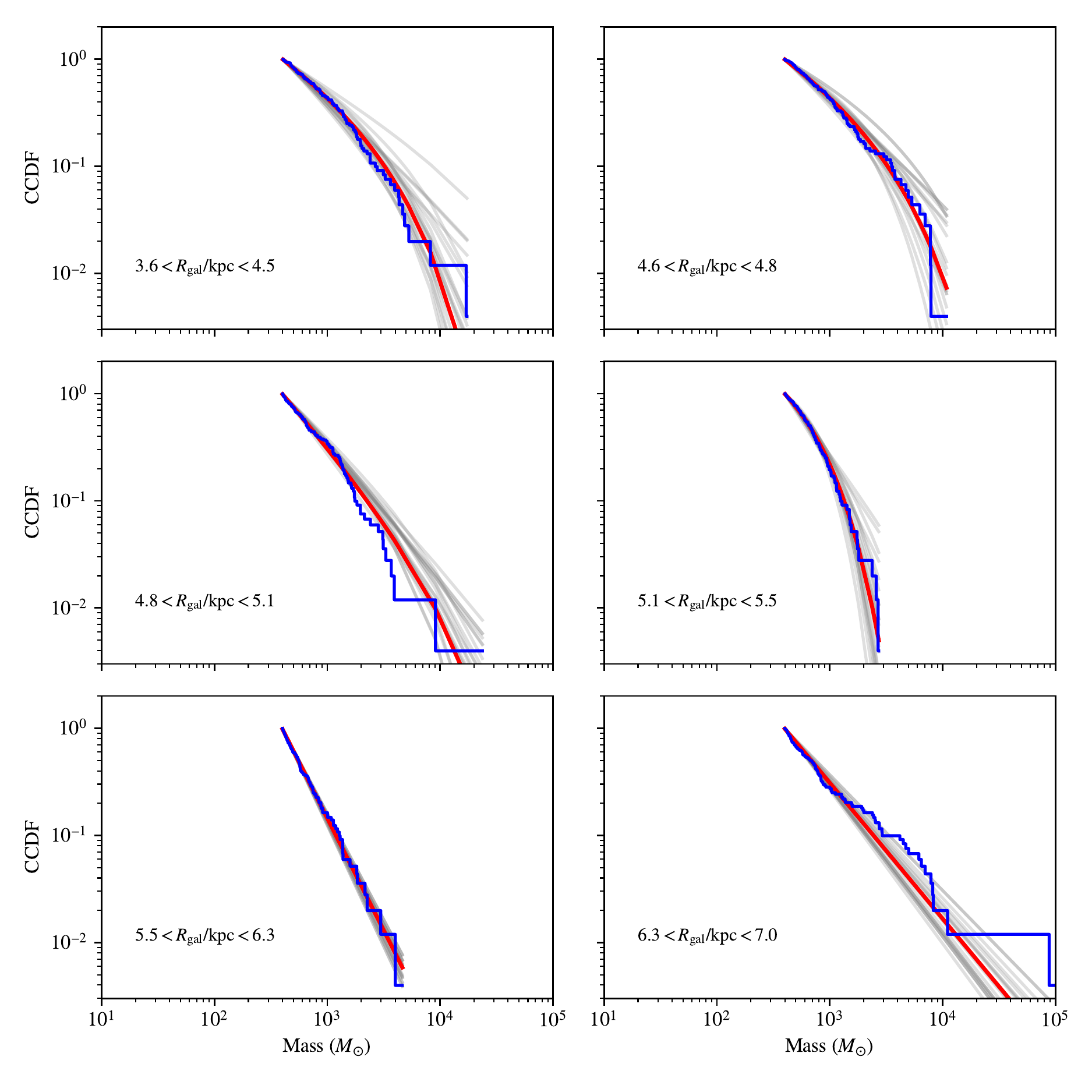}
    \caption{Mass Distributions for Clumps in Equal-Number bins of Galactocentric Radius.  The curves show the complementary cumulative mass distribution functions (CCDFs) for clump masses in six equal bins of 126 objects.  The red lines show power law fits to the CCDFs including truncations where they are detected, and the grey lines show 20 random draws from the posterior distribution of parameters.}
    \label{fig:mspec}
\end{figure*}

Table \ref{table:properties} summarizes the results of the mass distribution analysis for both the whole sample and individual radial bins. In particular, we report the maximum mass of clump found in the samples ($M_{\mathrm{max}}$) as well as the geometric mean of the five most massive clumps ($\langle M \rangle_5$). There is not a clear trend in either of these properties with $R_{\mathrm{gal}}$, mostly due to high mass structures associated with W51 in the last radial bin.  With the exception of these data, there is a modest trend toward lower mass structures at larger $R_{\mathrm{gal}}$.  We also report the indices for the two functional models, $\beta_{\mathrm{PL}}$ and $\beta_{\mathrm{TPL}}$ respectively.  For the TPL model, we report the characteristic mass of truncation and we compare the two models by reporting the peak log-probability associated with the most credible model in the fit ($\log_{10} p$).  This provides a discriminant for cases where one model provides a significantly better fit to the data than the other.  In the large $R_{\mathrm{gal}}$ bins, there is no good model from the truncated power-law fit and we only report values for the power-law model.  In most cases, both models provide good descriptions of the data and the clump mass distribution appears to have a power-law component with $\beta\sim 2$, though there is significant variation with $R_{\mathrm{gal}}$.

Figure \ref{fig:mspecall} shows the mass distribution for all the clumps in this sample and the power-law fits to the results. The power-law models provide excellent fits to the data with a consistent index between TPL and PL fits of $\beta \approx -2.25\pm 0.05$. There is very weak evidence for a truncation near $5\times 10^4~M_{\odot}$, but this model is not clearly favoured, and the simpler PL model provides an equally good model for the results.

Despite this clear aggregate behaviour over the survey, separating the samples into bins of $R_{\mathrm{gal}}$ shows significantly different mass distributions in the different bins (see Figure \ref{fig:mspec}). Notably, at small $R_{\mathrm{gal}}$ there is better evidence for a characteristic mass scale of $\sim6\,000~M_{\odot}$, though this result is marginal.  More strikingly, the mass distribution index for the PL fits shows a sharp decrease in the bin $5.5<R_{\mathrm{gal}}/\mathrm{kpc}<6.3$ where the index drops to $\beta_{\mathrm{PL}}=-3.1\pm 0.2$ and is significantly different from the other bins.  In contrast, the innermost bins also show the shallowest distributions $\beta_{\mathrm{PL}}>-2$. These variations are likely from sampling different parts of the galaxy. The steep mass distribution seems to arise because our mass distribution mostly samples the inter-arm regions of the galaxy at these radii.  The shallower slopes appear associated with the molecular ring structures and a richer molecular ISM. What remains clear is that aggregation of these different populations over a large piece of the galaxy produces a single power-law mass distribution with index of $\beta\approx -2.25$.  This emergent distribution reflects the typical gravitational fragmentation conditions over the inner galaxy, though there remains evidence that local conditions can produce significant variation.

\subsection{Dense Gas Distribution}

\begin{figure*}
\begin{center}
	\includegraphics[width=0.8\textwidth]{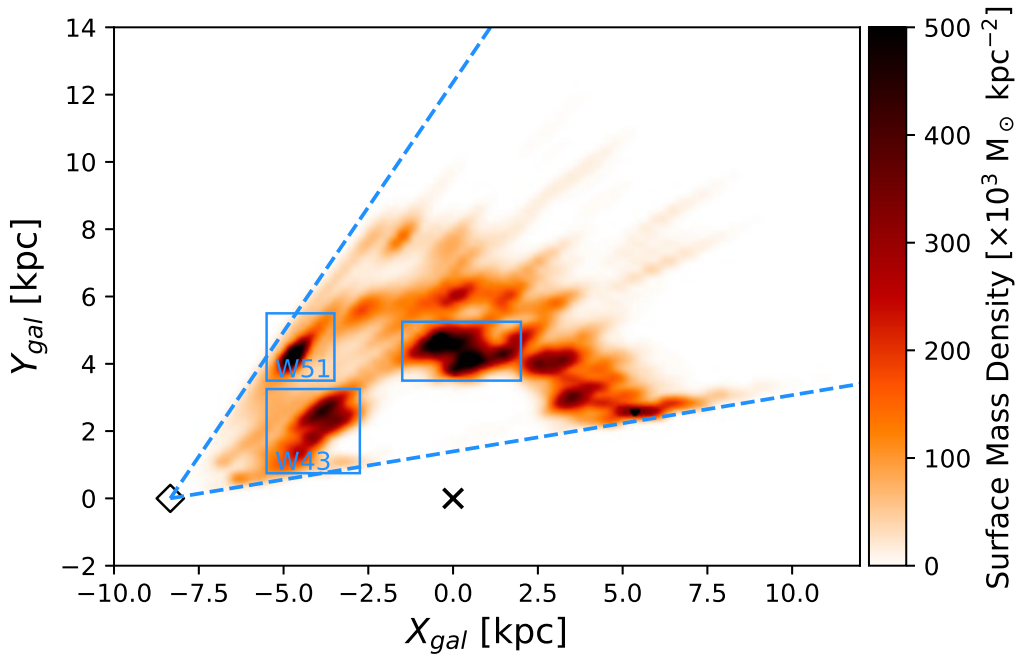}
\caption{Dense gas mass surface density over the Galactic disk. Each object is drawn from its DPDF $10^3$ times. For each draw, the mass of the clump is placed in a grid of 0.25 kpc pixels. The solar position and Galactic center are shown as a diamond and an X, respectively. The three major GMCs are enclosed by blue rectangles, and the dashed lines indicate the longitude limits of this survey.}
\label{fig:dg}
\end{center}
\end{figure*}

The dense gas mass surface density on the Galactic Plane is estimated by Monte Carlo sampling from the clump DPDFs. For each clump, we randomly draw from its DPDF $10^3$ times. Each time we calculate its mass from its observed flux density and simulated distance, and place that mass in the appropriate location in a grid of $0.25 \text{ kpc} \times 0.25 \text{ kpc}$ pixels representing the Galactic disk. The resulting map is shown in Figure \ref{fig:dg}. Spiral arm structure can be seen in two arcs roughly 4 kpc and 6 kpc from the Galactic center. Much of $R_\text{gal}\lesssim3$ kpc was excluded due to significant radial streaming motion corrupting the circular motions, keeping only those clumps with robust trigonometric parallax measurements \citep[see][]{BGPS8}. Sampling the entire DPDF results in some smearing effects along the line of sight, especially for sources more distant from the solar position, where the uncertainties tend to be larger, and the detected objects scarcer. We see little beyond $R_\text{gal}=8$ kpc. This is due to a combination of factors. On the near side of the Galactic center, our longitude range limits our observable $R_\text{gal}$ range. On the far side, we must first look through what is on the near side. Here our sensitivity at large heliocentric distances limits our ability to observe structures at large $R_\text{gal}$. As discussed in Section 1, at $d_\odot > 7$ kpc, we are only able to resolve entire clouds. On top of that, the amount of dense molecular gas drops off with increasing $R_\text{gal}$, so that where we can only resolve clouds, clouds are less likely to exist.

There is a marked, but unlabeled, region opposite the kinematic distance tangent point from the major GMC complex W43. It is possible that this region could be clumps which were mistakenly placed at the far kinematic distance, when they should be at the near distance within W43. There are many clumps with velocities putting them in one of these two regions, not all of which have had their kinematic distance ambiguity resolved. The Monte Carlo sampling of these unresolved DPDFs has contributed to the mirrored appearance of these two regions about the tangent point. However, this effect does not affect our conclusions about how clump properties vary with Galactocentric radius, nor about clump properties drawn from clumps with well-constrained distances, such as those in Figures 2, 3, 4, 5, and 6. There is also a significant number of clumps whose distances are well constrained in both regions, primarily due to 8 $\mu$m absorption features.

\begin{figure*}
\begin{center}
	\includegraphics[width=0.43\textwidth]{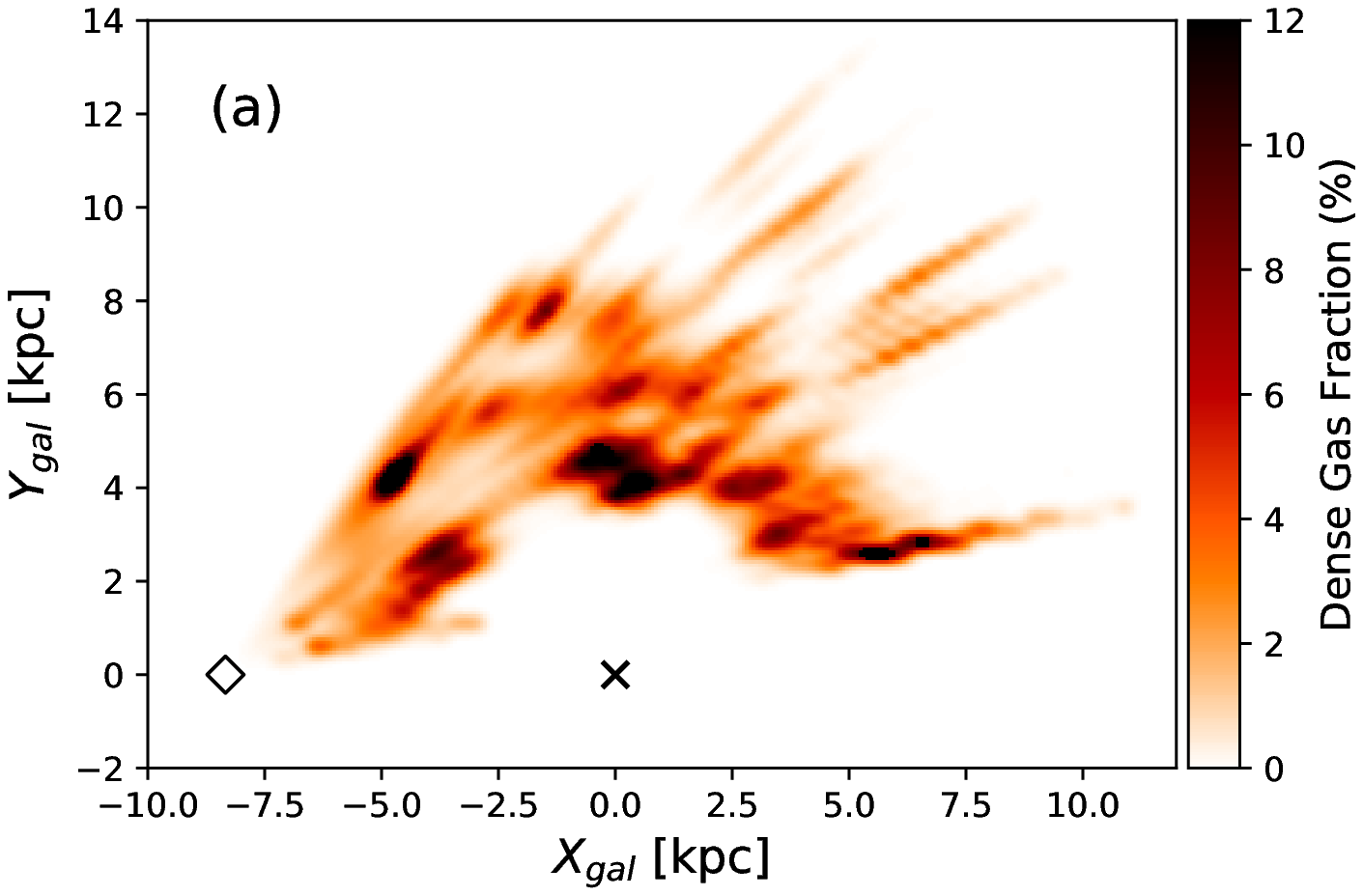}
	\includegraphics[width=0.36\textwidth]{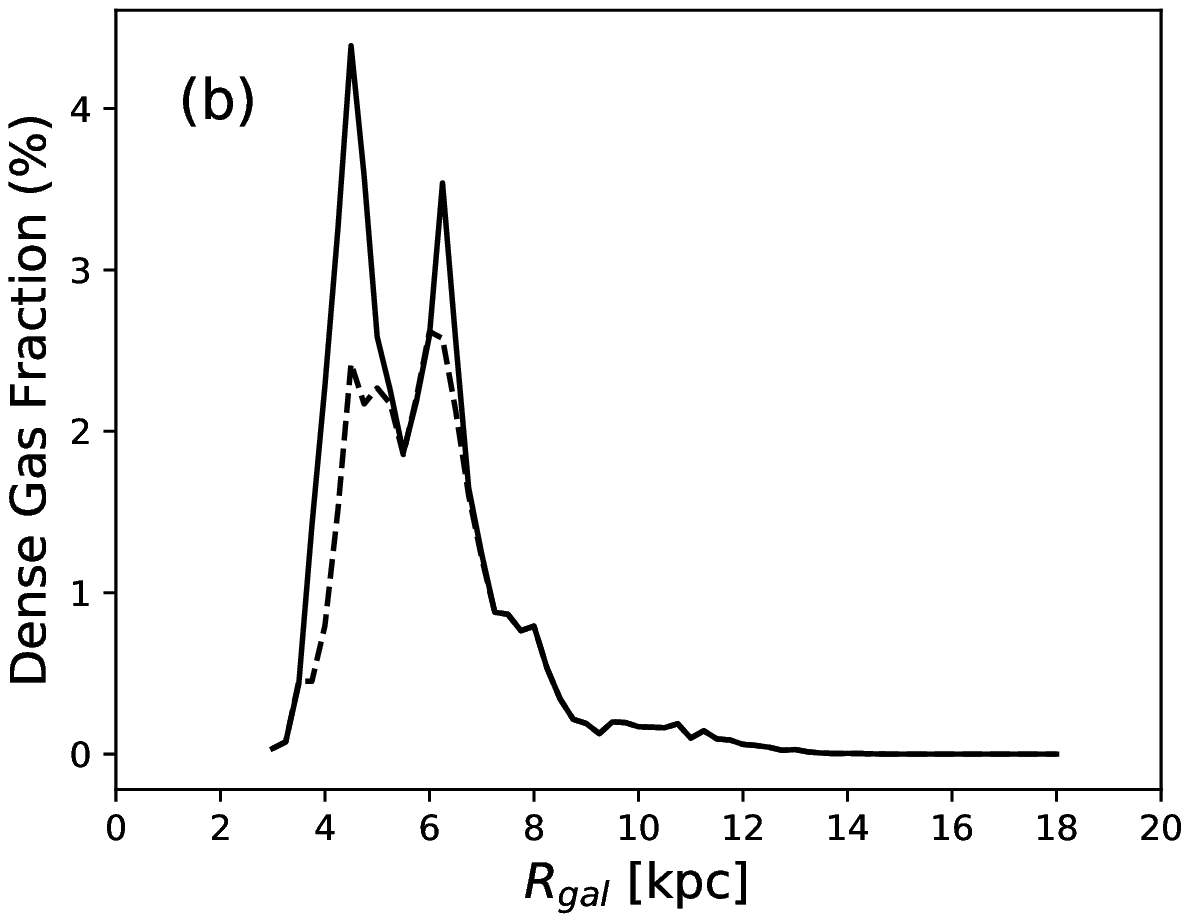}
	\includegraphics[width=0.43\textwidth]{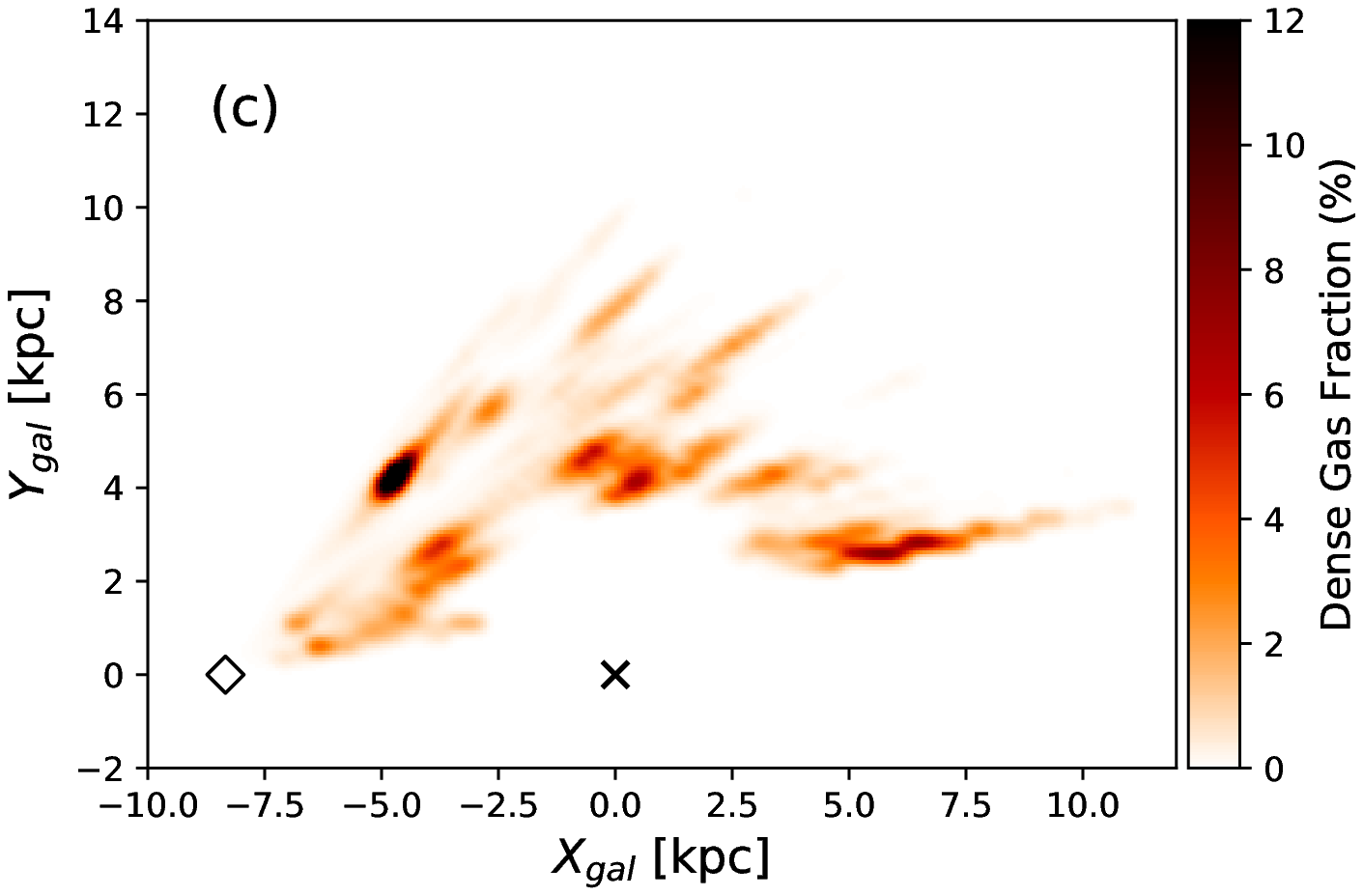}
	\includegraphics[width=0.36\textwidth]{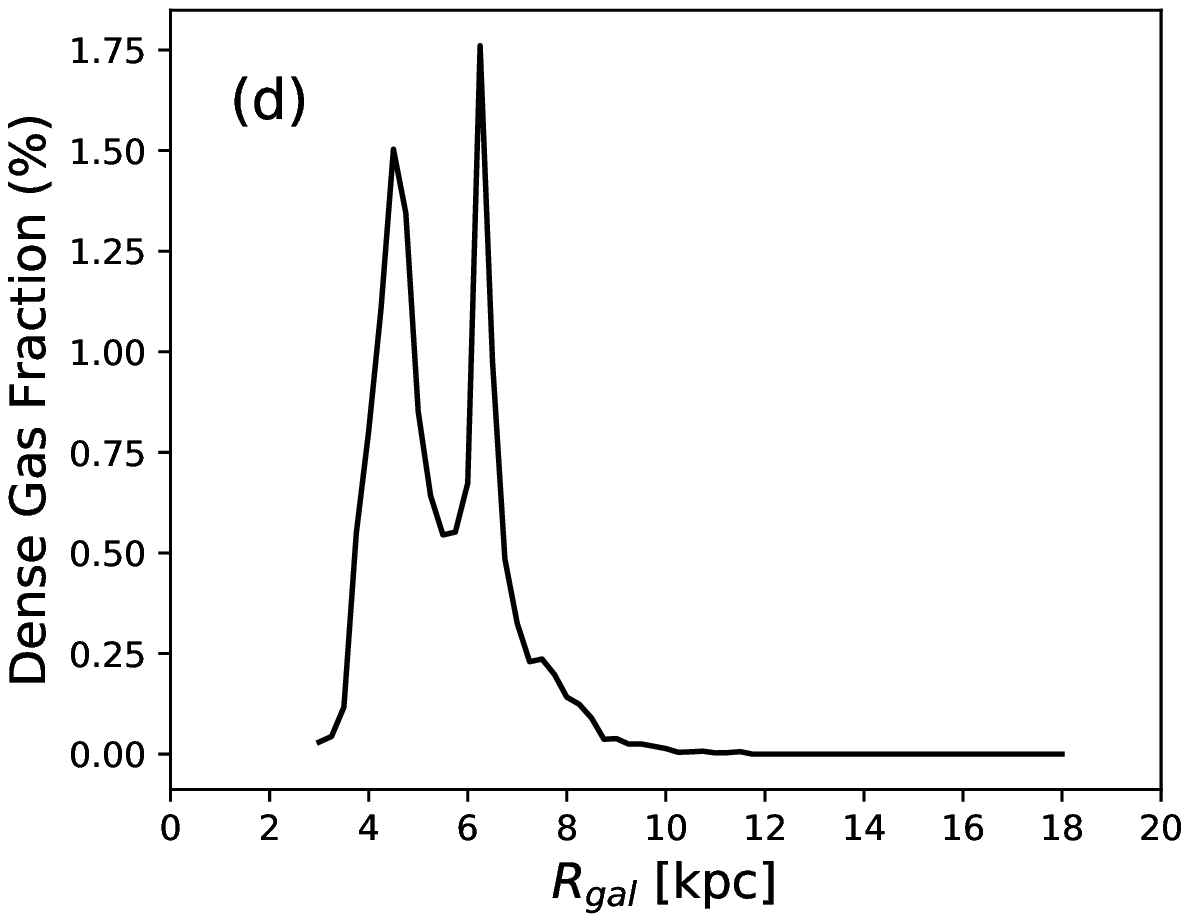}
\caption{(a) Map of the dense gas fraction in the Galactic plane, made by dividing the dense gas distribution as in Figure \ref{fig:dg} by the molecular hydrogen distribution from the model of \citet{Wolfire03}. The peak value is 47\%. (b) Dense gas fraction from (a) as a function of Galactocentric radius. The dashed line excludes the major GMCs indicated in Figure \ref{fig:dg}; the spikes at 4 kpc and 6 kpc remain, although they are less pronounced. (c) As in (a), but only including structures with $n \ge 10^3$ cm$^{-3}$. The peak value is 44\%. (d) Azimuthally averaged dense gas fraction from (c) as a function of Galactocentric radius. Note that the \citet{Wolfire03} model does not include enhancements for spiral arms, so the dense gas fraction in these spikes is likely somewhat lower in reality.}
\label{fig:dgf}
\end{center}
\end{figure*}

The azimuthally averaged H$_2$ model of \citet{Wolfire03} was used to convert the dense gas mass surface density map in Figure \ref{fig:dg} into a dense gas fraction map, seen in Figure \ref{fig:dgf} (a). Figure \ref{fig:dgf} (b) shows the dense gas fraction azimuthally averaged over the observed area. These panels show what may be spiral arms in arcs where the dense gas fraction is enhanced, and spikes corresponding to those arcs, respectively. As the map in Figure \ref{fig:dg} includes H$_2$ at densities hardly above those considered in the \citet{Wolfire03} model, we also include panels (c) and (d). These are the same as (a) and (b), respectively, but only considering clumps with number density $n \ge 10^3$ cm$^{-3}$. The spiral arms are again seen in the spikes around 4 kpc and 6 kpc. Note that the \citet{Wolfire03} model does not include enhancements for spiral arms, so the dense gas fraction in these spikes is likely somewhat lower in reality.

\section{Discussion}

\subsection{Mass Sensitivity}

The Hi-GAL survey had a raw gas mass sensitivity limit of 250 M$_\odot$ (5$\sigma$) at distances up to 20 kpc (corresponding to a 1$\sigma$ rms of 100 mJy at 250 $\mu$m, assuming a temperature of 20 K and a gas-to-dust conversion ratio of 100). However, because of source confusion, our actual sensitivity was about an order of magnitude worse. Figure \ref{fig:lognorms} indicates that we can see clouds with masses of 3000 M$_\odot$ out to distances of 14 kpc. This decreased sensitivity impacts what we can detect on the far side of the Galactic center, as discussed in Section 4.3. While, according to Figure \ref{fig:MnRgal}, there should exist clumps of mass 100 M$_\odot$ at $R_\text{gal}=6$ kpc, it is evident in Figure \ref{fig:malmquist} (bottom) and Figure \ref{fig:lognorms} (top) that we do not see these clumps because of confusion.

\subsection{The Effects of GMCs}

Considering Figure \ref{fig:dg}, it appears possible that the trend in normalized mass seen in Figure \ref{fig:MnRgal} and the structures seen in Figure \ref{fig:dgf} b and d may be dominated by the labeled GMCs. To test for the dominance of these GMCs, we removed the sources in the boxed regions from our analysis. The slope of the plot of normalized mass against Galactocentric radius is still negative when these sources are removed, and indeed slightly steeper. The steeper slope is most likely due to the removal of W51, which contributes high-mass objects at relatively large $R_\text{gal}$ values. Since we are dealing with normalized, not true, masses, the slope is tracing the declining value of the typical flux density with Galactocentric radius.

Figure \ref{fig:dgf}b also shows the radial dense gas fraction when the major GMCs are excluded.The two spikes around 4 kpc and 6 kpc are still present, although not as pronounced, due to the most massive regions being removed. This suggests that the decrease in dense gas and dense gas fraction with Galactocentric radius is not simply a result of a few massive GMC complexes.

\begin{figure}
    \centering
    \includegraphics[width=\columnwidth]{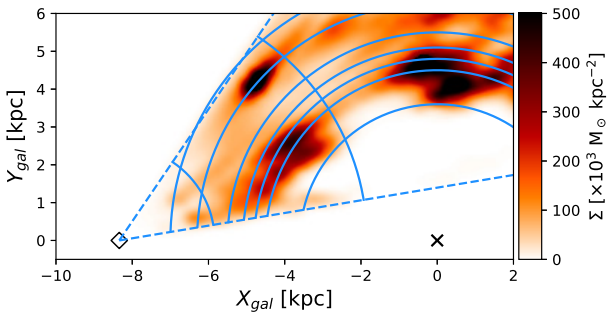}
    \caption{Dense gas mass fraction map annotated with radial bins used in mass spectrum analysis. Heliocentric arcs indicate the limits of the minimum-bias slice. This figure illustrates how the bins used in the mass distribution analysis relate to the dense gas structures seen in the Galaxy.  The effects of the W51 region will be seen in the outer two radial bins.}
    \label{fig:DGbins}
\end{figure}

The mass distribution analysis shows steeper mass distributions between 5.1 and 6.3 kpc.  Figure \ref{fig:DGbins} shows the locations of these radial bins in our selected survey region.  These two bins are notable since they do not contain a significant fraction of the GMC structure in this region.  There is a small portion of the W51 complex contained in this region but it does not appear to be enough to change the structure of the mass distributions.  While this analysis lacks sufficient numbers of regions in the two categories, it is suggestive that regions with the large GMC complexes also show relatively top-heavy mass distributions for their global clump populations. 

\subsection{Comparison to Literature}

\subsubsection{Galactocentric Radius Trends}

Our work is in general agreement with previous work on the trends of GMCs with Galactocentric radius using different tracers of dense gas. \citet{Scoville87} found that GMCs peak in the molecular ring, located at $R_\text{gal} = 4-7$ kpc. While we now consider the molecular ring to be nearer the Galactic center and this range to include spiral arms, we do see a higher dense gas fraction in that range of Galactocentric radii (Figure \ref{fig:dgf}). Later, \citet{Bronfman00} found that the number of GMCs and $L_\text{FIR}$ peak at $R_\text{gal} = 4-5$ kpc. Indeed, we find that --- while we exclude much that is nearer the Galactic center than 4 kpc, due to the limitations of kinematic distance determinations --- the typical mass of cloud clumps decreases moving outward from 4 kpc. The analysis of normalized masses suggests that this behaviour persists with a minimally biased sample.

Comparing to extragalactic work, where face-on views make for much easier mapping of GMCs, we also see agreement.  On kpc-scales in galaxies, the brightness of CO typically peaks in the centres or in rings at small galactocentric radius \citep{Leroy09,Utomo17}.  The result holds at higher linear resolution where \citet{Gratier12} found that CO brightness decreases with $R_\text{gal}$ in M33, reflecting the same trend we see in Figures \ref{fig:MnRgal} and \ref{fig:MRgal}.  Mass distribution analysis in \citet[][Braine et al., submitted]{Rosolowsky07} shows that the characteristic mass of GMCs peaks at intermediate values of $R_\text{gal}=2-4$ kpc in M33 (noting the smaller size of the galaxy). As with our comparison to the Milky Way work of \citet{Bronfman00} above, we exclude sources inside the 4 kpc ring for kinematic reasons, and therefore cannot conclude where the characteristic mass peaks, but we can conclude it falls off outside of 4 kpc, which is an intermediate $R_\text{gal}$ value.

There are several possibilities for the cause of such an $M-R_\text{gal}$ relation, the simplest explanation being that the decrease in gas surface density in the outer Galaxy means that clouds will accrue mass more slowly, leading to less massive clouds further from the Galactic center. \citet{Blitz06} argued that hydrostatic pressure drives the conversion of H\textsc{i} into H$_2$, which is less efficient at larger radii. \citet{Meidt16} argued that the mass of dense gas cores is directly related to the cloud surface density, a property which is inherited from the environment. This would also lead to lower masses at larger radii.


\citet{Rosolowsky07} additionally found that the mass distribution for GMCs in M33 is not different in the spiral arms vs. in the interarm regions. Contrarily, \citet{Colombo14} found that, in M51, the mass distribution varies not by $R_\text{gal}$, but between arm and interarm regions, noting that the arm-to-interarm ratio decreases with $R_\text{gal}$. Unfortunately, while we do see interarm gas, we do not have enough of the Galactic plane mapped to be definitive about the distribution of clumps within the interarm regions. For this reason, it is necessary to analyze the Southern Hemisphere and thus complete a map of the Galactic plane.

\subsubsection{Dense Gas Fraction}

We calculated dense gas fraction in the same manner as \citet{BGPS13} using the BGPS data. With Hi-GAL data, we find significantly higher dense gas mass fractions as compared with BGPS. We find that half of all non-zero pixels have $>$1\% dense gas mass fraction, and that 10\% of the non-zero pixels are $>$5\% dense gas. Only a quarter of pixels above each limit remain when only structures with $n \ge 10^3$ cm$^{-3}$ are considered. This compares with the finding by \citet{BGPS13} that while there are pixels with $>$5\% dense gas mass fraction, the vast majority of pixels have $<$1\% dense gas. While we see higher dense gas fractions, due to our increased sensitivity, our results remain in agreement.
\citet{Battisti14} compared the masses found in BGPS sources to the masses of the parent GMCs in GRS, finding a mean fraction of $0.11^{+0.12}_{-0.06}$. Our work is consistent with this, finding only three pixels above their $1\sigma$ errors.

\section{Conclusions}

We have mapped the physical distribution of the dense molecular cloud clumps found in Hi-GAL within the Galactic longitude range $10^\circ < \ell < 56^\circ$. This range corresponds to the extent of the Galactic Ring Survey, which provided the majority of our line-of-sight velocities. These maps were made from 10\,124 clumps which have, at minimum, a kinematic distance. Of these, 5\,405 clumps had their distances well constrained through Bayesian techniques. Maps of the dense gas and dense gas fraction show features which are suggestive of spiral arms.

While the Malmquist bias impairs our ability to detect trends in clump mass and radius as a function of Galactocentric radius, we have analyzed these quantities over restricted heliocentric distances. This revealed that the mean clump mass decreases with distance from the Galactic center, while the mean clump radius stays constant, indicating an additional decrease in mean clump density. Previous work concerning the distribution of clump masses, both Galactic and extragalactic, are in general agreement with our findings.

\section*{Acknowledgements}
SPIRE has been developed by a consortium of institutes led by Cardiff University (UK) and including Univ. Lethbridge (Canada); NAOC (China); CEA, LAM (France); IFSI, Univ. Padua (Italy); IAC (Spain); Stockholm Observatory (Sweden); Imperial College London, RAL, UCL-MSSL, UKATC, Univ. Sussex (UK); and Caltech, JPL, NHSC, Univ. Colorado (USA). This development has been supported by national funding agencies: CSA (Canada); NAOC (China); CEA, CNES, CNRS (France); ASI (Italy); MCINN (Spain); SNSB (Sweden); STFC (UK); and NASA (USA).

The authors wish to thank John Bally and Sergio Molinari for helpful discussions. ER is supported by a Discovery Grant from NSERC of Canada.  This work made use of the {\sc astropy} \citep{astropy} and {\sc matplotlib} \citep{matplotlib} software packages and we are grateful for the continued efforts of the open source community at improving our capabilities in science.

Part of this work based on archival data, software or online services provided by the ASI SCIENCE DATA CENTER (ASDC).

\textit{Facility}: Herschel (SPIRE)

\bibliographystyle{mnras}
\bibliography{Bibliography}{}

\end{document}